\documentclass[12pt,preprint]{aastex}

\shorttitle{PARTICLE ACCELERATION PROCESSES IN THE COMA CLUSTER}
\shortauthors{KUO, HWANG, \& IP}

\begin{document}

\title{Diagnostic Signatures of Radio and HXR Emission on Particle Acceleration
Processes in the Coma Cluster}

\author{Ping-Hung Kuo, Chorng-Yuan Hwang, and Wing-Huen Ip}
\affil{Institute of Astronomy, National Central University, Chung-Li 32054, Taiwan}
\email{d882001@astro.ncu.edu.tw, hwangcy@astro.ncu.edu.tw, wingip@astro.ncu.edu.tw}
\begin{abstract}

We investigate theoretical models for the radio halo and hard X-ray (HXR)
excess in the Coma galaxy cluster. Time-independent and time-dependent
re-acceleration models for relativistic electrons have been carried out to
study the formation of the radio halo and HXR excess. In these models, the
relativistic electrons are injected by merger shocks and re-accelerated by
ensuing violent turbulence. The effects of different Mach numbers of the merger
shocks on the radio and HXR excess emission are also investigated. We adopt 6
$\mu$G as the central magnetic field and reproduce the observed radio spectra
via the synchrotron emission. We also obtain a central ``plateau" in the radio
spectral-index distribution, which have been observed in radio emission
distribution. Our models can also produce the observed HXR excess emission via
the inverse Compton scattering of the cosmic microwave background photons. We
find that only the merger shocks with the Mach numbers around 1.6--2 can
produce results in agreement with both the radio and HXR emission in the Coma
cluster.

\end{abstract}


\section{INTRODUCTION}

Radio halos in galaxy clusters are diffuse radio sources possessing large sizes
and steep spectra. The radio emission must be produced by the synchrotron
radiation of relativistic electrons. Nonetheless, the sources of these
relativistic electrons are still unclear. In the intracluster medium (ICM),
relativistic electrons lose energy on the time scale of order $\sim10^{8}$
years because of inverse Compton and synchrotron losses \citep[e.g.,][]{ip99}.
Because of the short lifetimes of the relativistic electrons, it is difficult
to interpret the large size of radio halos as the result of the diffusion of
the relativistic electrons injected from radio galaxies \citep{jaf77}.
Consequently, for the formation of radio halos, a significant level of
re-acceleration might be involved. The secondary electron model first proposed
by \citet{den80} provides a different scenario for the origin of the radio halo
and can avoid the problem of re-acceleration for the relativistic electrons in
the primary electron models. However, this model encounters serious problems
when comparing with observations \citep[for a recent review, see][]{bru02}.

Coma C in the Coma cluster is the prototype of radio halos in galaxy clusters.
Being the best studied example, Coma C has been observed at many different
radio wavebands \citep*[e.g.,][]{sch87,kim90,ven90,gio93,dei97,thi03}.
\citet{dei97} argued that the measurement at 2.7 GHz by \citet{sch87} is too
low and suggested that the integrated spectrum might have no tendency to
steepen as suggested by \citet{sch87}; however, this strong steepening of the
radio spectrum at high frequencies was confirmed by \citet{thi03}.
\citet{gio93} used the radio data at 1380 MHz \citep{kim90} and 326 MHz
\citep{ven90} to derive the radial distribution of the spectral index in Coma
C, and they found a central ``plateau" with a size of $\sim15\arcmin$ for the
spectral-index distribution. In the central region, the value of the spectral
index is $\sim0.8$; in the outside region, the spectral index strongly steepens
as the radius increases.

Observations with \textit{BeppoSAX} and the \textit{Rossi X-ray Timing
Explorer} (\textit{RXTE}) have detected a hard X-ray (HXR) excess with respect
to thermal emission from the Coma cluster \citep*{fus99,rep99,rep02}. The HXR
excesses from several other clusters have also been reported
\citep{kaa99,fus00,fus01,gru02}. The most favored mechanism of the HXR excess
is inverse Compton scattering (ICS) of the cosmic microwave background (CMB)
photons by relativistic electrons. Since these electrons with energy
$\gamma\sim10^{4}$ will also produce radio synchrotron radiation, the HXR
excesses and radio halos may originate from the same electron population. In
the Coma cluster, the volume-averaged values of magnetic fields deduced from
the comparison of the radio with the HXR excess emission is $\sim$ 0.1--0.3
$\mu$G \citep{fus99,rep99,rep02}. These low values of field strength are not
consistent with those deduced from the measurements of Faraday rotation.
\citet*{cla01} have shown that many clusters have relatively large ($\sim$ 4--8
$\mu$G) fields. For the Coma cluster, \citet{kim90} found a central field
strength of $1.7\pm0.9\ \mu$G, and \citet{fer95} estimated the strength to be
$6\pm1\ \mu$G. It is thus very important to see whether it is possible to
produce the observed HXR excess via the inverse Compton mechanism with such a
high central magnetic field. Nonetheless, some uncertainties in both methods
may lead to the discrepancy \citep*{new02}. An alternative interpretation of
the HXR excess is non-thermal bremsstrahlung from supra-thermal electrons
\citep*{ens99,bla00,dog00,sar00}. However, a huge amount of energy is necessary
in this model to produce the observed HXR excess \citep{pet01,bla00}.

\citet{bru01} proposed a two-phase model to interpret the radial steepening of
the spectral-index distribution in Coma C. In this two-phase model, the
relativistic electrons were injected during a first phase in the past and
re-accelerated during a second phase up to present time. In such a
re-acceleration model, there must be a cutoff in the electron spectrum because
of the balance between the loss and the gain of the electron energy. A cutoff
in the electron spectrum should lead to a cutoff in the emissivity of
synchrotron radiation and then in the integrated radio spectrum. If the
magnetic fields have a radial-decrease profile, the cutoff frequencies will
decrease with the increasing radius. Consequently, a radial steepening of the
spectral index between two fixed frequencies will appear. \citet{bru01}
successfully reproduced the radial steepening of the spectral index, the radio
spectrum steepening at high frequencies, and the HXR excess in the Coma
cluster; however, the central ``plateau" in the spectral-index distribution was
not well explained. The authors adopted a central field strength of $\lesssim
3\ \mu$G in their models. As pointed in their work, the spectral index would be
very steep in low magnetic fields and flat in moderate ones. It would be
important to know whether a stronger central magnetic field can reproduce the
central ``plateau" in the spectral-index distribution without violating the
observational constraints from the radio and HXR emission.

Cluster mergers are very violent events and release a large amount of energy
($\sim10^{64}$ ergs). Merger shocks and violent turbulence must play an
important role in the generation and re-acceleration of relativistic electrons
\citep{sar01}. Nonetheless, \citet{gab03} claimed that the shocks generated by
major mergers, mergers between clusters with comparable mass, is too weak to
account for the spectral slopes of the non-thermal emission. The Mach numbers
of the shocks in major mergers are of order of unity in their simulations and
roughly consistent with the Mach number $\sim$ 2 observed in Cygnus A
\citep*{mar99}. However, if a significant level of re-acceleration is involved,
the evolved spectra of relativistic electrons may be able to account for the
observed spectra of non-thermal emission.

In this paper, we investigate the radio and HXR excess emission in the Coma
cluster assuming the magnetic fields possessing a central field strength of $6\
\mu$G and a radial-decrease profile. The effects of different Mach numbers of
merger shocks on the formation of radio halos and HXR excesses are also
investigated. We assume that relativistic electrons are injected by merger
shocks and re-accelerated by ensuing violent turbulence. The turbulence is
expected to be more violent at the moment right after the merger shocks and
then gradually decays with time \citep{ric01}; therefore, particles in the ICM
might obtain stronger re-acceleration in the early period of mergers and then,
following the decay of turbulence, the strength of re-acceleration would
decrease. In this work, both time-independent and time-dependent
re-acceleration models are considered. The radial variation of the spectral
index of synchrotron radiation in the Coma cluster is computed to test the
scenarios of the cosmic-ray electron re-acceleration. In particular, we try to
obtain a central ``plateau" in the spectral-index distribution that is in
agreement with observational results reported by \citet{gio93}. We also
investigate the production of an HXR excess with the assumption that the HXR
excess is due to ICS of the CMB photons by the cosmic-ray electron distribution
that we obtained from the spectral index fitting. Comparing the modelling
results with the observations, we find that the Mach numbers of the merger
shocks have to be in a very small range to form the radio halo and the HXR
excess observed in the Coma cluster.

In this paper, $H_{0}=50$ km s$^{-1}$ Mpc$^{-1}$ is assumed. The red shift $z$
of the Coma cluster is $\sim0.0233$, so that the distance is $\sim$ 140 Mpc and
$1\arcmin$ corresponds to $\sim$ 40 kpc. The virial radius of 3.28 Mpc
\citep{gir98} is adopted as the radius of the Coma cluster.

\section{PARTICLE ACCELERATION MODELS}\label{sPAM}

The energy variation of particles can be expressed as:
\begin{equation}\label{Eloss}
  -\frac{d\gamma}{dt}=b_{0}+b_{1}\gamma+b_{2}\gamma^{2},
\end{equation}
where $b_{0}=b_{\mathrm{Coul}}$, $b_{1}=b_{\mathrm{brem}}-b_{\mathrm{acc}}$,
and $b_{2}=b_{\mathrm{syn}}+b_{\mathrm{IC}}$. The coefficients of the loss
rates due to the Coulomb and bremsstrahlung losses can be approximated as
\citep{sar99}:
\[ b_{\mathrm{Coul}}\approx1.2\times 10^{-12}n_{gas}\left[1.0+\frac{\ln(\gamma/n_{gas})}{75}\right]\ \mbox{s}^{-1}, \]
and
\[ b_{\mathrm{brem}}\approx1.51\times 10^{-16}n_{gas}[\ln(\gamma)+0.36]\ \mbox{s}^{-1}, \]
where $n_{gas}$ is the gas density. For the Coma cluster, it can be defined as:
\begin{equation}\label{Egas}
  n_{gas}(r)=n_{0}f_{gas}(r)=n_{0}[1+(\frac{r}{r_{gas}})^2]^{-3\beta_{gas}/2},
\end{equation}
where $n_{0}=2.89\times10^{-3}$ cm$^{-3}$, $r_{gas}=10\farcm5 \approx 0.42$
Mpc, and $\beta_{gas}=0.75$ \citep*{bri92}. We set $\gamma=10^{3}$ and
$n_{gas}=10^{-3}$ cm$^{-3}$ into the logarithms for simplicity.

The coefficients in the loss function of inverse Compton scattering and
synchrotron radiation can be expressed as \citep{sar99}:
\[ b_{\mathrm{IC}}=1.37\times 10^{-20}(1+\mathrm{z})^{4}\ \mbox{s}^{-1}, \]
and
\[ b_{\mathrm{syn}}=1.3\times 10^{-21}\left(\frac{B}{1\ \mu G}\right)^{2}\ \mbox{s}^{-1}, \]
where z $\ll 1$ is assumed in these calculations.

To account for the temporal dependence of the electron re-acceleration due to
violent turbulence, the general form of the acceleration function
$b_{\mathrm{acc}}$ is given by:
\begin{equation}\label{Ebacc}
  b_{\mathrm{acc}}(r,t)=b_{a}(r)g(t),
\end{equation}
where
\begin{equation}\label{Eba0}
  b_{a}(r)=a_{0}+a_{1}f_{a}(r)
\end{equation}
\begin{equation}\label{Etimef}
g(t) = 1+Ae^{-Dt}.
\end{equation}
The parameters $A$ and $D$ are the amplification factor and the dissipation
factor, respectively. The cases with $A=D=0$ are called time-independent
acceleration, and the cases with $A\neq0$ and $D\neq0$ are called
time-dependent acceleration. According to the numerical work of \citet{ric01},
the maximum temperature of mergers is $\lesssim$ three times of the initial
temperature before merging and returns to that when the passed times after
merging are greater than 1 Gyr. Because the sources for heating gas are
turbulence and shocks generated by merging, it is reasonable to relate the
temperature variation with the turbulence strength, i.e., the strength of
re-acceleration. Thus we choose the values of the amplification factor $A$ and
the dissipation factor $D$ under the constraint that the maximum value of
$g(t)$, i.e., $1+A$, is $\lesssim$ 3 and the value of $g(t)$ is $\sim$ 1 when
$t =$ 1 Gyr in our models.

The re-acceleration model in the two-phase model proposed by \citet{bru01} is a
time-independent acceleration model in our classification. \citet{bru01}
suggested that the re-acceleration function ($\chi(r)$ in their eq.\,[16]
corresponding to $b_{a}(r)$ here) can be parameterized as the sum of a uniform
large-scale component and a small-scale component and assumed that the uniform
large-scale component is caused by shocks and/or turbulence, possibly generated
during a recent merger, and the small-scale component assumed to be
proportional to the inverse of the typical distance between galaxies is due to
the amplification of these shocks/turbulence by the motion of the massive
galaxies in the cluster core. Following \citet{bru01}, we assume that the
acceleration component $a_{0}$ in the re-acceleration function $b_{a}(r)$ is
due to turbulence and spatially uniform in the cluster. The lower-hybrid-wave
turbulence proposed by \citet{eil99} is a possible scenario for this kind of
re-acceleration. They showed that large amplitude Alfv\'{e}n waves do generate
lower hybrid waves which collapse to produce localized, intense wave packets,
and even a modest level of lower hybrid turbulence can be very effective at
accelerating relativistic particles. For the $a_{1}f_{a}(r)$ component
(corresponding to the small-scale component in the two-phase model), we assume
that this is caused by the orbiting motion of galaxies. According to the work
of \citet{dei96}, the excited turbulent motions of ICM are related to the
galaxy density in the cluster and vary as $v_{turb}^{2}\propto n_{gal}$, so
that we assume $f_{a}(r)\propto n_{gal}$. The $\beta$-model is adopted for the
galaxy distribution in the Coma cluster:
\begin{equation}\label{Egal}
  n_{gal}(r) \propto f_{gal}(r)=[1+(\frac{r}{r_{gal}})^2]^{-3\beta_{gal}/2},
\end{equation}
where $r_{gal}\approx0.18$ Mpc and $\beta_{gal}=0.86$ \citep{gir98}. We define
$f_{a}(r)=f_{gal}(r)$.

For time-independent acceleration, the analytic solution of equation
(\ref{Eloss}) with the condition that $\gamma$ and $n_{gas}$ are constant in
the logarithms of the Coulomb and bremsstrahlung losses are (see, e.g.,
\citet{bru01}):
\begin{equation}\label{Eengt1}
  \tau=\frac{2}{\sqrt{|\eta|}}\left(\tanh^{-1} y-\tanh^{-1} y_{0}\right),
\end{equation}
where
\[ \tau=t-t_{0}, \]
\[ \eta=4b_{0}b_{2}-b_{1}^{2}, \]
\[ y=\frac{2b_{2}\gamma+b_{1}}{\sqrt{|\eta|}}, \]
\[ y_{0}=\frac{2b_{2}\gamma_{0}+b_{1}}{\sqrt{|\eta|}}. \]
The notation $\gamma_{0}$ is the energy of electrons at time $t_{0}$.

This solution can be rewritten as:
\begin{equation}\label{Eengt2}
  \gamma(t)=\frac{\sqrt{|\eta|}}{2b_{2}}\left(\frac{y_{0}+\tanh x}{1+y_{0}\tanh
  x}\right)-\frac{b_{1}}{2b_{2}},
\end{equation}
where
\[ x=\frac{\sqrt{|\eta|}}{2}\tau. \]
The cutoff energy is defined by setting $\gamma_{0}=\infty$ at time $t_{0}$ in
equation~(\ref{Eengt2}) and can be expressed as:
\begin{equation}\label{Ecutoff}
\gamma_{c}(t)=\frac{\sqrt{|\eta|}}{2b_{2}\tanh x}-\frac{b_{1}}{2b_{2}}.
\end{equation}

\section{INJECTION AND EVOLUTION OF THE ELECTRON SPECTRUM}\label{sINJ}

In our models, the relativistic electrons are assumed to be injected by merger
shocks. A power-law spectrum of relativistic electrons is expected from diffuse
shock acceleration. We assume that the injected spectrum has the form:
\begin{equation}\label{Einj}
  n_{e}(\gamma,r)=f_{e}(r)K_{e}\gamma^{-s},
\end{equation}
where $f_{e}(r)$ is assumed to be proportional to the gas distribution and is
normalized to be equal to $f_{gas}(r)$ as described in equation (\ref{Egas}),
i.e., $f_{e}(r)=f_{gas}(r)$. The parameter $K_{e}$ can be determined by
normalizing the theoretical radio spectrum to the observed data. The power-law
index $s$ is related to the Mach number $\mathcal{M}$ of the merger shocks by
the expression: $s=2(\mathcal{M}^{2}+1)/(\mathcal{M}^{2}-1)$
\citep[e.g.,][]{gab03}. For studying the effects of the Mach numbers of merger
shocks on the non-thermal emission in galaxy clusters, we adopt different
values for the power-law index $s$: 2.5, 3.3, 4.0, and 4.7 corresponding to
Mach numbers: 3, 2, 1.73, and 1.58, respectively. Without loss of generality,
we ignore the initial variation of the power-law index in the cluster for
simplicity.

Cosmic-ray protons can also generate secondary electrons. Nonetheless, it is
still unclear about the contribution of the secondary electrons to the
relativistic electrons in galaxy clusters. In our calculation, we consider only
the electrons injected by merger shocks, and assume that the secondary
electrons are negligible \citep*{kuo03}.

The evolution of the electron population is described by the kinetic equation:
\begin{equation}\label{Edif}
  \frac{\partial n_{e}(\gamma)}{\partial
t}=\frac{\partial}{\partial\gamma}[b(\gamma)n_{e}(\gamma)]+q(\gamma),
\end{equation}
where $b(\gamma)=b_{0}+b_{1}\gamma+b_{2}\gamma^{2}$. Here we assume no
continuous injections, i.e., $q(\gamma)=0$. The analytic solution of equation
(\ref{Edif}) for time-independent acceleration is then:
\begin{equation}\label{Edifs}
  n_{e}(\gamma,t)=n_{e}(\gamma_{0},t_{0})\left[\frac{1-\tanh^{2}x}{(1-y\tanh x)^{2}}\right],
\end{equation}
where $x$ and $y$ are defined in equations (\ref{Eengt2}) and (\ref{Eengt1}),
respectively.

\section{DISTRIBUTION OF MAGNETIC FIELDS IN COMA}\label{sMF}

\begin{figure}
  \plotone{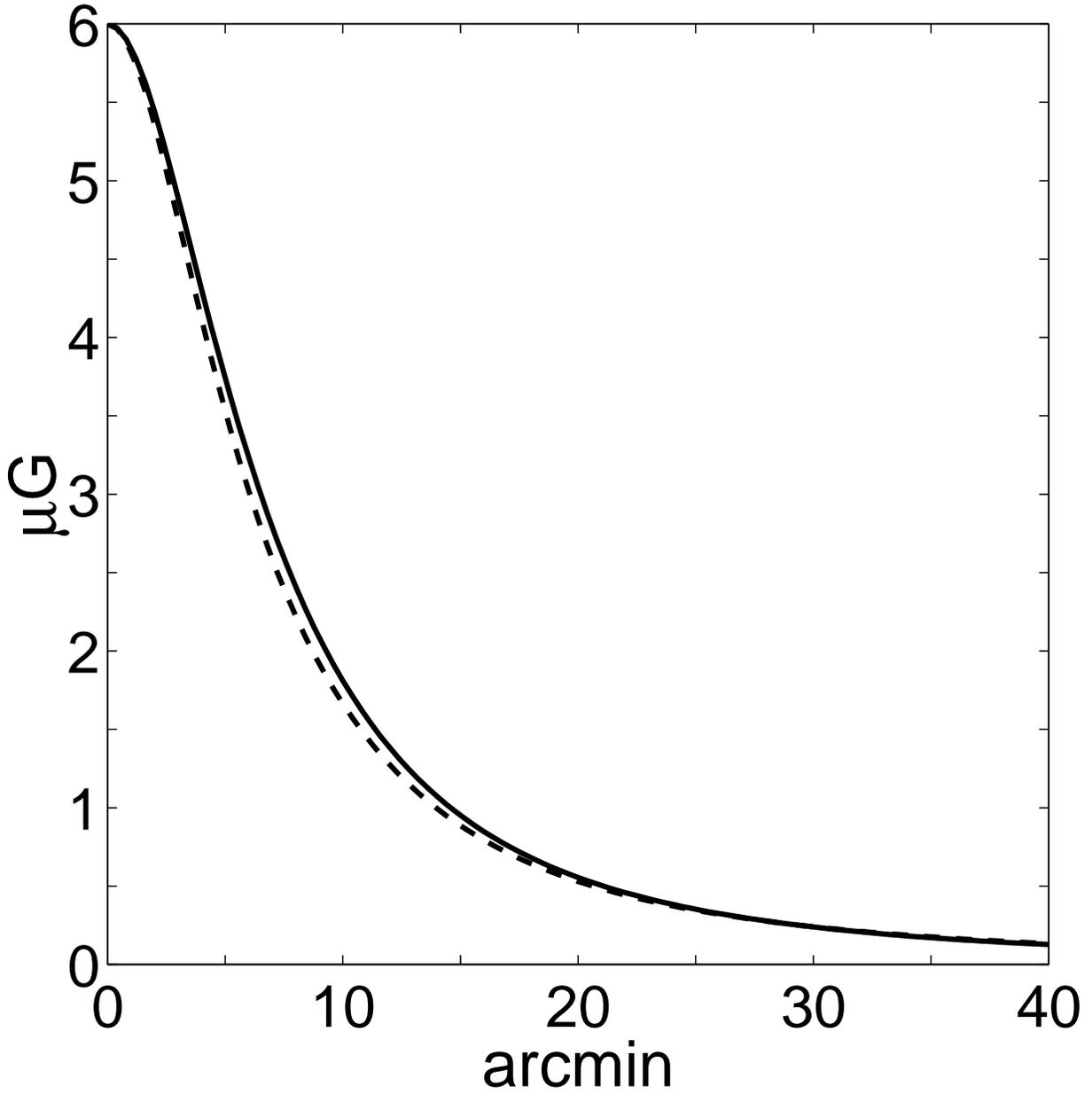}
  \caption{Profiles of magnetic fields. The profiles with $(m,n)=(0.7,0.3)$ (\textit{solid
  curve}), and $(m,n)=(0.5,0.4)$ (\textit{dashed curve}) are shown. The profile with
  $(m,n)=(0.5,0.4)$ was proposed by \citet{jaf80}.
  }\label{fig1}
\end{figure}

Using the techniques of dimensional analysis for studying turbulent structures,
\citet{jaf80} derived a magnetic field model determining by the gas and galaxy
distributions and described by:
\begin{equation}\label{EBf}
  B(r)=B_{0}f_{B}(r)=B_{0}[f_{gas}(r)]^{m}[f_{gal}(r)]^{n},
\end{equation}
where $f_{gas}(r)$ and $f_{gal}(r)$ are defined in equations (\ref{Egas}) and
(\ref{Egal}), and $(m,n)=(0.5,0.4)$. For the Coma cluster, there are other
constraints that can be used to determine the profile of magnetic fields; the
observed spectral-index distribution possesses a central ``plateau" and
strongly steepens outside this region. It is expected that the profile of the
magnetic field strongly affected the size of the central ``plateau" and the
steepness of the spectral-index distribution outside this region. We assume the
central field strength $B_{0}$ is 6 $\mu$G. The profile with $(m,n)=(0.7,0.3)$
shown in Figure~\ref{fig1} is chosen in our paper in order to make the size of
the central ``plateau" and the steepness of the spectral-index distribution
outside this region calculated in the acceleration models agreeing with the
observations.

The orbiting motion of galaxies might amplify the seed fields to grow into the
present magnetic field in galaxy clusters \citep*{jaf80,rol81,ruz89}. However,
\citet{dey92} has shown that the turbulent dynamo driven by the galaxy motion
is difficult to produce the present micro-gauss fields. Cluster merging, a very
energetic process, may offer a possibility to produce the observed fields.
Because the decay time of the magnetic field is very long, the magnetic field
will gradually build up under successive merging \citep{tri93}. It is
reasonable to assume that the observed fields at the present time is the
amplified fields after the last merging. This is the reason for the assumption
that the magnetic field is time-invariant in our models.

We note that the profile of $(m,n)=(0.7,0.3)$ is adopted for the
purpose of reproducing the observed spectral-index distribution
and might be different from a real distribution of cluster
magnetic fields. For examples, the profiles of simulated magnetic
fields are flat in the core region and the fields are $\sim$ 1
$\mu$G \citep*{dol02}. From the Faraday rotation measurements, the
magnetic fields are $\sim$ 4--8 $\mu$G out to $\sim$ 0.75 Mpc from
cluster centers \citep{cla01}. The adopted central magnetic field
is higher than the simulated results but lower than the
observational ones.

\section{MODELLING PROCEDURE}\label{sCA}

In this section, we briefly describe the methods for calculations. We assume
the electron spectra have a power law at time $t= 0$, which corresponds to the
moment right after the main-merger shock. The electron spectra have an initial
power-law index $s$ according to the Mach number of the shock and would then
evolve according to the formulism described in $\S$~\ref{sPAM} and
$\S$~\ref{sINJ}. Since a flat central region in the spectral-index
distributions of the radio-halo emission may exist only in a short period
during the evolution of radio halos, we calculate the brightness distributions
at two fixed frequencies, 326 MHz and 1380 MHz, for different evolution time
scales and derive the spectral-index distribution from them. To fit and compare
our models with observations, we first vary the acceleration function
$b_{\mathrm{acc}}(r,t)$ and judge whether the corresponding spectral-index
distributions are consistent with the observations according to two criteria:
(1) there is a flat central region with $\alpha^{1380}_{326}=$ 0.8 as reported
by \citet{gio93} and (2) $\alpha^{1380}_{326}=$ 1.8 is located in the range of
16$\arcmin$--18$\arcmin$ to agree with the results obtained by \citet{dei97}.
The spectral-index distributions at different time $t$ with a step of 0.2 Gyr
are computed until consistency with the criteria have been obtained. Second, we
also calculate the spatial brightness distributions at 326 MHz to compare with
the observed one at 90 cm \citep[kindly provided by F. Govoni]{gov01}. Third,
we normalize the theoretical radio spectrum to the observed data at 430 MHz and
then determine the normalized parameter $K_{e}$ in equation~(\ref{Einj}).
Finally, the HXR spectrum is calculated in a region with a projected radius of
50$\arcmin$ and compared with the observational data from \textit{BeppoSAX}
\citep{fus99} and \textit{OSSE} \citep*{rep94}. The thermal temperature of the
cluster is assumed to be 8.21 keV \citep{hug93} in our calculations.

\section{MODEL RESULTS FOR COMA C}

\subsection{Time-Independent Re-Acceleration}

We study the time-independent re-acceleration models by assuming that the
amplification and dissipation factors in the acceleration function are zero.
Different initial power-law indices corresponding to different Mach numbers are
considered. The values of the parameters in the time-independent models are
listed in Table~\ref{tbl}.

\begin{figure}
  \plotone{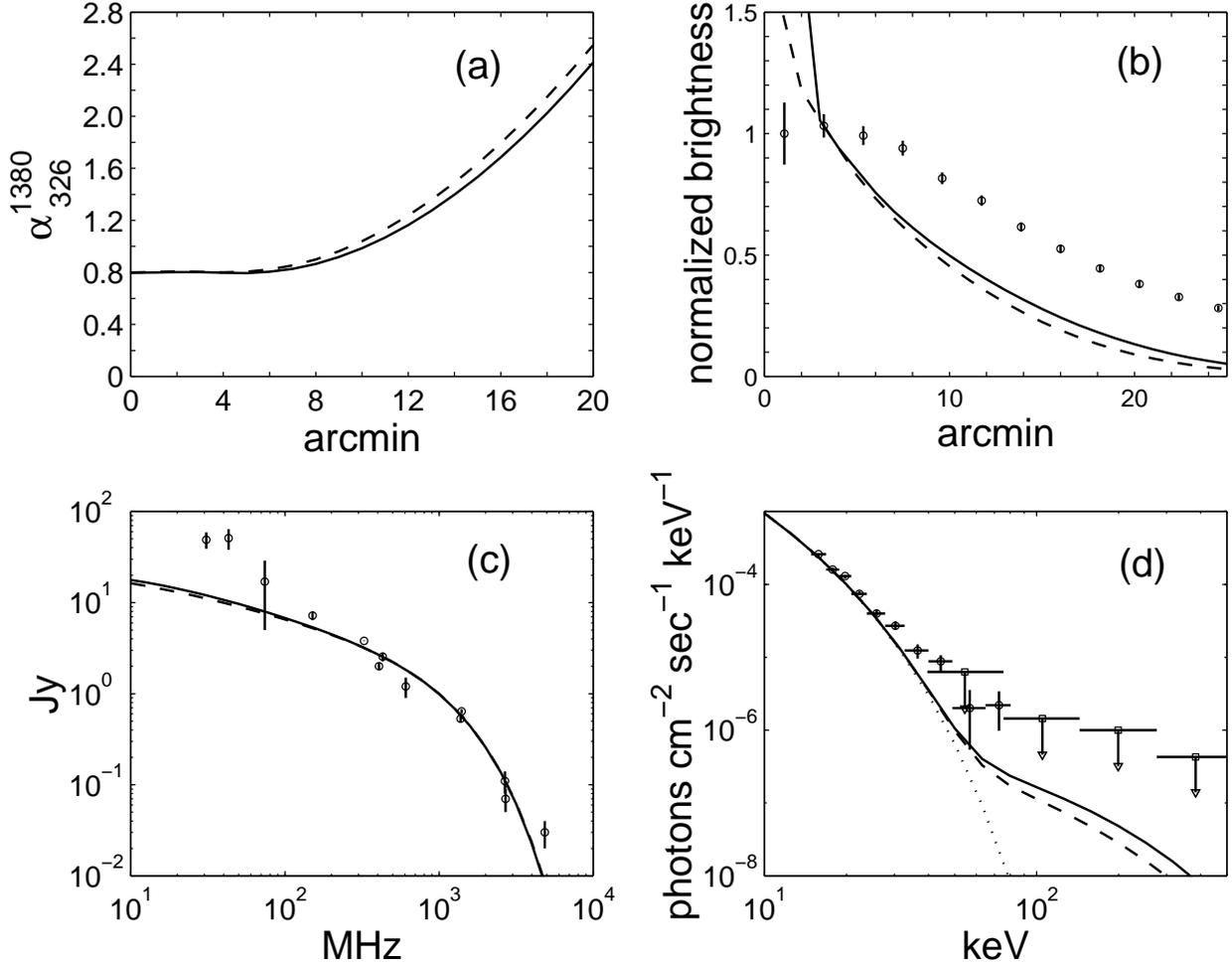}
  \caption{Time-independent models with the power-law index $s=$ 2.5, models A1 (\textit{dashed
  curve}) and A2 (\textit{solid curve}). (a) Spectral-index distributions. (b) Brightness
  distributions. The observed brightness is taken from \citet{gov01}. (c) Integrated radio
  spectra. The measured data (\textit{circle}) are taken from \citet{thi03}. (d) HXR spectra. The \textit{dotted curve} is a
  thermal model with $kT=$ 8.21 keV. The \textit{circle} points are the data from the \textit{BeppoSAX} measurement
  \citep{fus99} and the \textit{square} points are the upper limits from \textit{OSSE} \citep{rep94}.
  }\label{fig2}
\end{figure}

Figure~\ref{fig2} shows the results from the models with the power-law index
$s=$ 2.5 (models A1 and A2). The radii of the flat central regions in the
spectral-index distributions are $\approx 6\arcmin$ in both models, which is
roughly in consistent with the observations. Nonetheless, the spatial
brightness distributions are much more concentrated toward the central regions
comparing with observations; the integrated radio spectra also deviate from the
measured data both at the low and high frequency parts. The HXR spectra are too
low to compare with the observations. These results show that a merger shock
with a Mach number $\mathcal{M} \approx 3$ cannot produce the radio and HXR
emission observed in the Coma cluster.

\begin{figure}
  \plotone{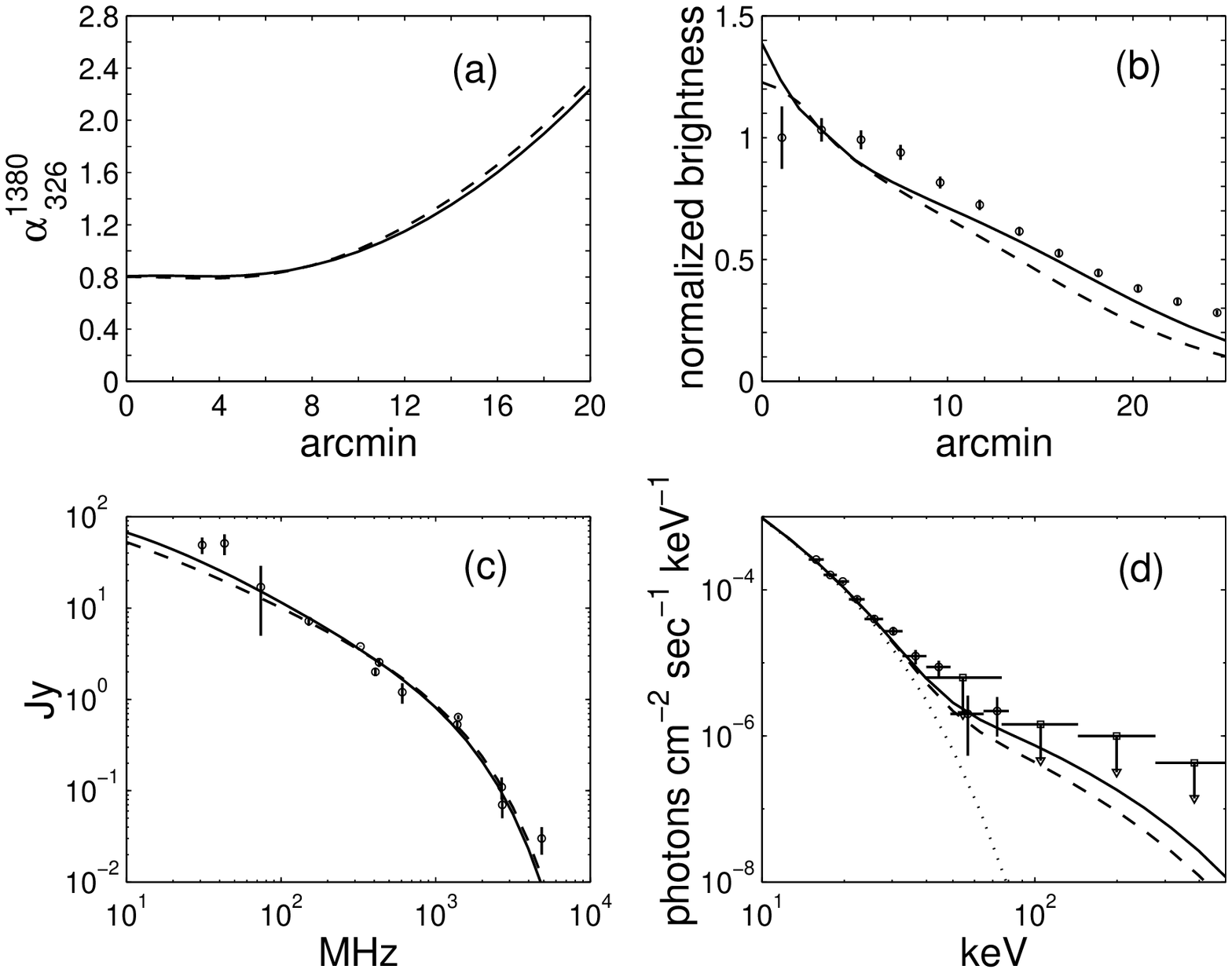}
  \caption{Time-independent models with the power-law index $s=$ 3.3, models A3 (\textit{dashed
  curve}) and A4 (\textit{solid curve}). (a) Spectral-index distributions. (b) Brightness
  distributions. (c) Integrated radio spectra. (d) HXR spectra.
  }\label{fig3}
\end{figure}

\begin{figure}
  \plotone{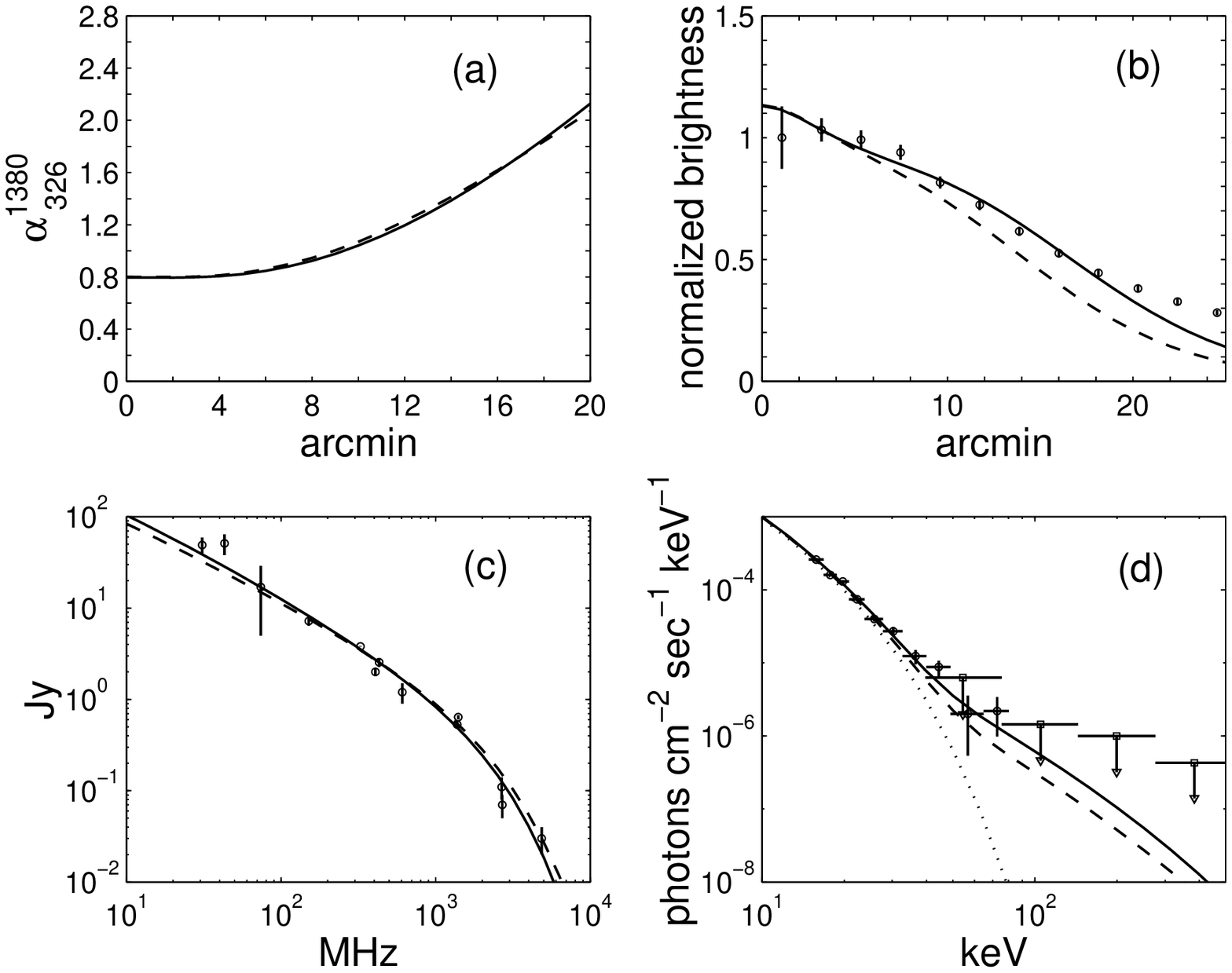}
  \caption{Time-independent models with the power-law index $s=$ 4.0, models A5 (\textit{dashed
  curve}) and A6 (\textit{solid curve}). (a) Spectral-index distributions. (b) Brightness
  distributions. (c) Integrated radio spectra. (d) HXR spectra.
  }\label{fig4}
\end{figure}

Figure~\ref{fig3} and Figure~\ref{fig4} show the results with the power-law
index $s= 3.3$ (models A3 and A4) and the power-law index $s= 4$ (models A5 and
A6), respectively. All these models can produce a flat spectral-index
distribution with a radius of $\approx 5\arcmin$--6$\arcmin$ in the central
regions. The brightness distributions and the HXR spectra are all roughly
consistent with the observations in these models, although distributions seem
to be better described by the $s=4$ models than the $s=3.3$ ones. These models
can also match the integrated radio spectra very well; the spectra also seem to
be better fitted by the models with $s=4$ than those with $s=3.3$. These
results indicate that merger shocks with Mach numbers $\mathcal{M} \lesssim$ 2
can produce the main features of the radio and HXR emission observed in the
Coma cluster.

\begin{figure}
  \plotone{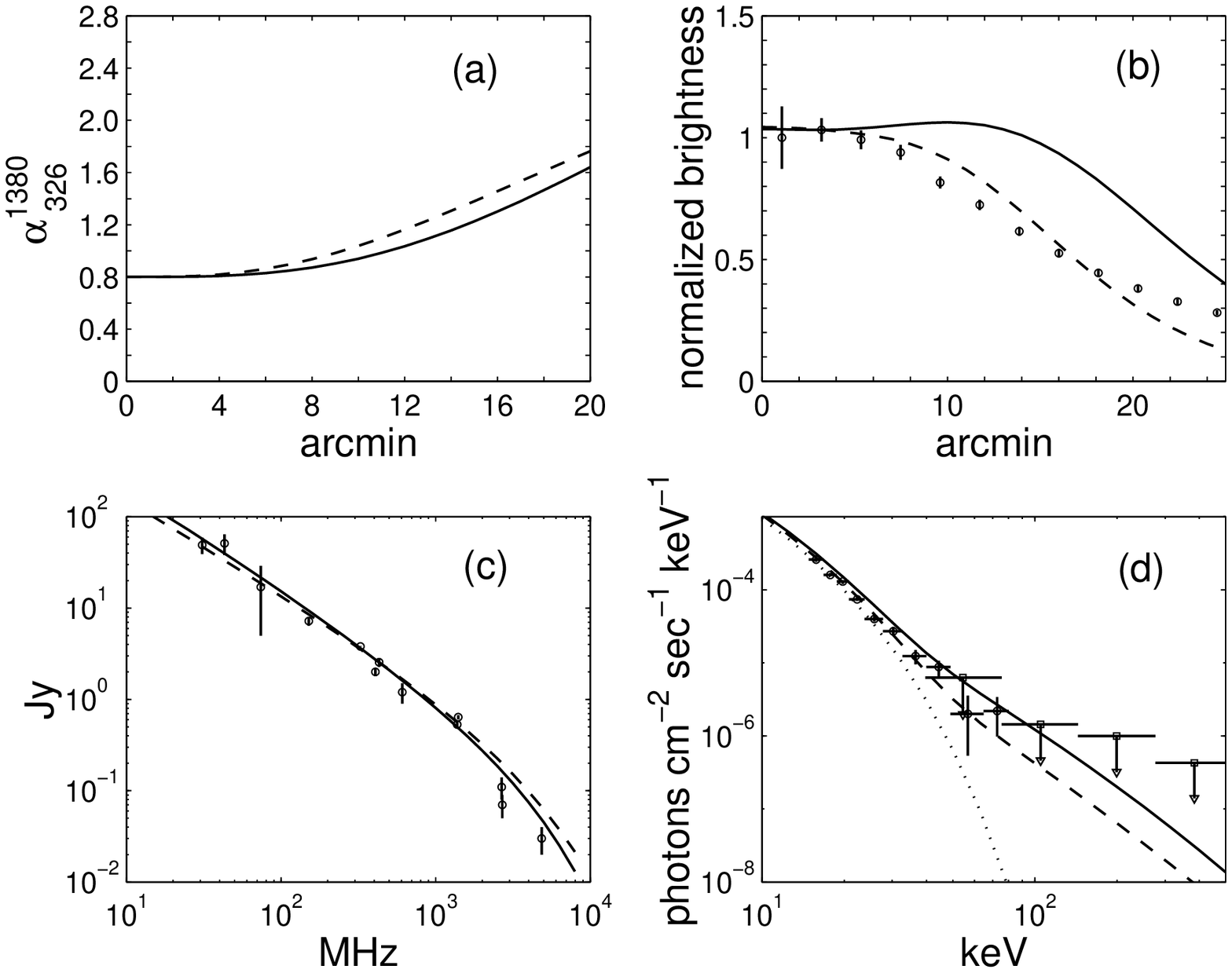}
  \caption{Time-independent models with the power-law index $s=$ 4.7, models A7 (\textit{dashed
  curve}) and A8 (\textit{solid curve}). (a) Spectral-index distributions. (b) Brightness
  distributions. (c) Integrated radio spectra. (d) HXR spectra.
  }\label{fig5}
\end{figure}

Figure~\ref{fig5} shows the results with the power-law index $s= 4.7$ (models
A7 and A8). We cannot obtain the spectral-index distributions that have a
spectral index $\alpha^{1380}_{326}=$ 1.8 in the region
16$\arcmin$--18$\arcmin$ that satisfy the second criterion described in
$\S$~\ref{sCA} for models with a power-law index $s= 4.7$ or higher when we fit
the central flat spectral index to $\alpha^{1380}_{326} \approx$ 0.8.
Nonetheless, we try to find a flat central distribution in the spectral index
distribution for these models by ignoring the second criterion. The results are
shown in Figure~\ref{fig5}. The radii of the flat central regions are $\approx
4\arcmin$ in model A7 and $\approx 6\arcmin$ in model A8. The spectral index
$\alpha^{1380}_{326}=$ 1.8 locates at radius larger than $20\arcmin$ in both
models. The spatial brightness distributions show a flat distribution in the
central regions; this is particularly obvious for models A8. Both models do not
fit the integrated radio spectral data at high frequencies; model A8 also
overproduces the HXR excess. These results indicate that it is difficult to
produce all observed features of the radio and HXR emission in the Coma cluster
by mergers shocks with Mach numbers $\mathcal{M} \leq 1.6$.

According to the results shown in Figures~\ref{fig2}--\ref{fig5}, a central
``plateau" in the spectral-index distribution can be produced for all these
models. Nonetheless, only the models with the initial electron power-law index
$4.7 > s \gtrsim 3.3$ can produce results that match all observed features of
the radio and HXR emission in the Coma cluster; this corresponds to a very
narrow range for the Mach numbers of the merger shocks, i.e., $1.6 <
\mathcal{M} \lesssim 2$.

\subsection{Time-Dependent Re-Acceleration}

In this section, we show the effects of time-dependent re-acceleration by
assuming that the amplification factor $A=1$ and the dissipation factor
$D=2$~Gyr$^{-1}$. We note that the values of $A$ and $D$ are constrained by the
conditions $1+A \lesssim 3$ and $Ae^{-Dt}$ becoming sufficient small as $t$
approaching 1 Gyr. Since we just want to investigate the effects of the
time-dependent re-acceleration on our current models, we fix the values of $A$
and $D$ for simplicity. A suitable but different choice of $A$ and $D$ could
affect the results, particularly, on the evolving time scales that fit the
observations. The values of the parameters in the time-dependent models are
listed in Table~\ref{tb2}. Since the results from the time-independent models
with $s=2.5$ and 4.7 strongly deviate from the observations, we consider only
the cases of $s=3.3$ and 4 in the time-dependent models.

\begin{figure}
  \plotone{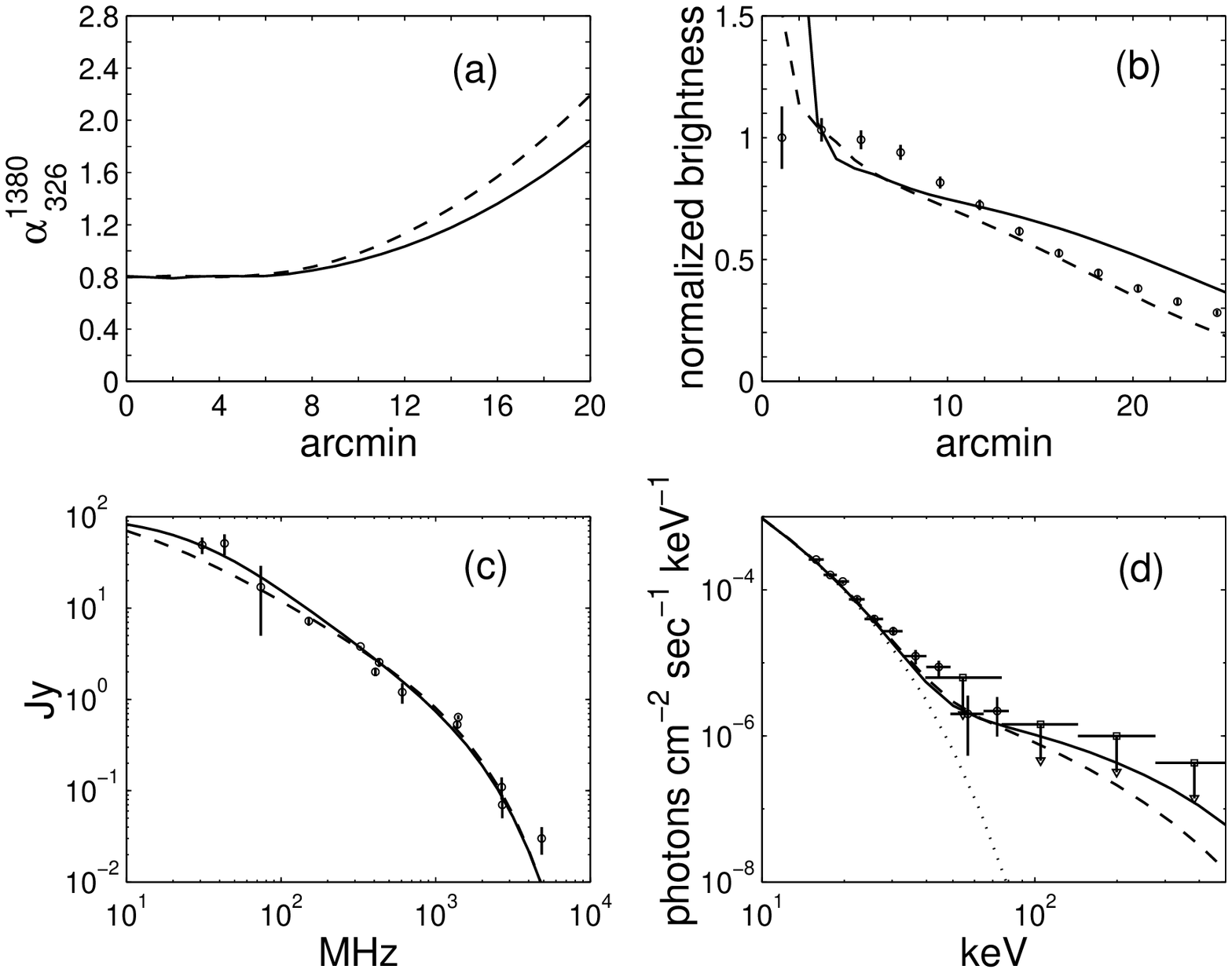}
  \caption{Time-dependent models with the power-law index $s=$ 3.3, models B1 (\textit{dashed
  curve}) and B2 (\textit{solid curve}). (a) Spectral-index distributions. (b) Brightness
  distributions. (c) Integrated radio spectra. (d) HXR spectra.
  }\label{fig6}
\end{figure}

Figure~\ref{fig6} shows the results with the power-law index $s=3.3$ (models B1
and B2). The flat central regions in the spectral-index distributions have
radii of $\approx 6\arcmin$ in both models B1 and B2. However, we note that in
the model B2 the spectral index $\alpha^{1380}_{326}=$ 1.8 locates at $\approx
20\arcmin$, which does not satisfy the second criterion stated in
$\S$~\ref{sCA}. The brightness distributions are concentrated to the central
regions resembling those from models A1 and A2. These results indicate that
models with time-dependent re-acceleration have effects similar to those with
flatter initial electron spectra. This also shows that merger shocks with Mach
numbers, $\mathcal{M} \geq 2$, i.e., $s \leq 3.3$, cannot produce the radio
emission observed in the Coma cluster.

\begin{figure}
  \plotone{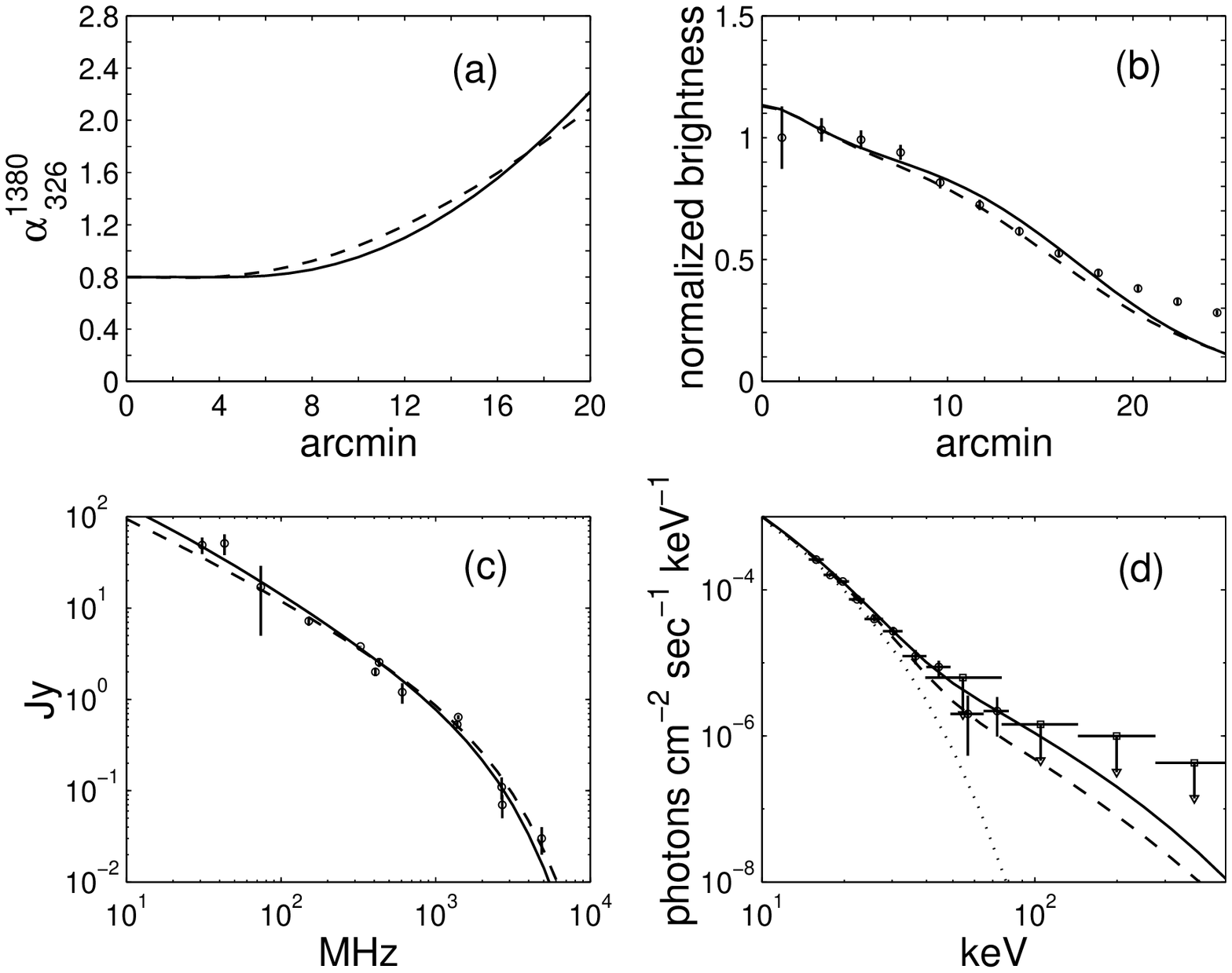}
  \caption{Time-dependent models with the power-law index $s=$ 4.0, models B3 (\textit{dashed
  curve}) and B4 (\textit{solid curve}). (a) Spectral-index distributions. (b) Brightness
  distributions. (c) Integrated radio spectra. (d) HXR spectra.
  }\label{fig7}
\end{figure}

Figure~\ref{fig7} shows the results with the power-law index $s=4$
(models B3 and B4). The flat central regions in the spectral-index
distributions have a radius of $\approx 4\arcmin$ in model B3 and
$\approx 6\arcmin$ in model B4. The integrated radio spectra and
the HXR spectra obtained from both models are all in very good
agreement with the observations. The brightness distributions also
agree very well with the observations in inner part of the
cluster, but are slightly lower than the observed ones at the
radii greater than $18\arcmin$.

We find that the radio spatial brightness distribution is very sensitive in
these time-dependent models. Our results indicate that the models with
time-dependent re-acceleration have effects on the brightness distribution
similar to those with flatter initial electron spectra. These effects can be
understood because the time-dependent re-acceleration function provides a
larger re-acceleration to the electrons at the early stage of the evolution;
this is similar to have a flatter initial electron spectrum. Nonetheless,
time-dependent models can still fit the integrated radio spectra and the HXR
excess emission better. Combining the results from the time-independent and
time-dependent models, we find that only merger shocks with Mach numbers, $1.6
< \mathcal{M} < 2$, i.e., the power-law index $4.7 > s > 3.3$, can reproduce
the results agreeing with the observations in the Coma cluster.

\section{DISCUSSION}

\begin{figure}
  \plotone{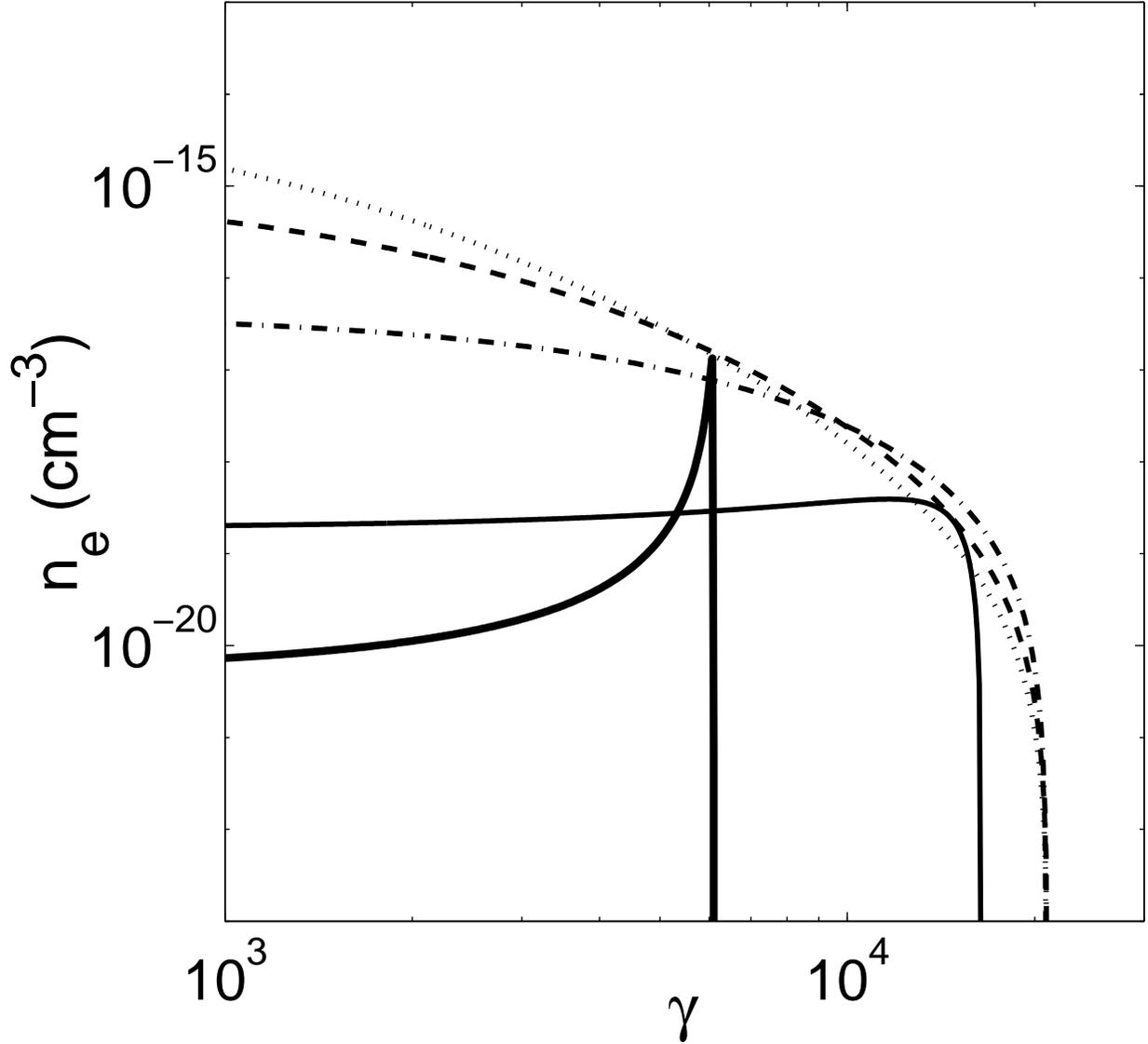}
  \caption{Electron spectra at different radii. The parameters are from model A6. The spectra at
  $r=$ 0 (\textit{thick solid curve}), $10\arcmin$ (\textit{solid curve}), $30\arcmin$
  (\textit{dash-dotted curve}), $50\arcmin$ (\textit{dashed curve}), and $70\arcmin$ (\textit{dotted
  curve}) are shown.
  }\label{fig8}
\end{figure}

\begin{figure}
  \plotone{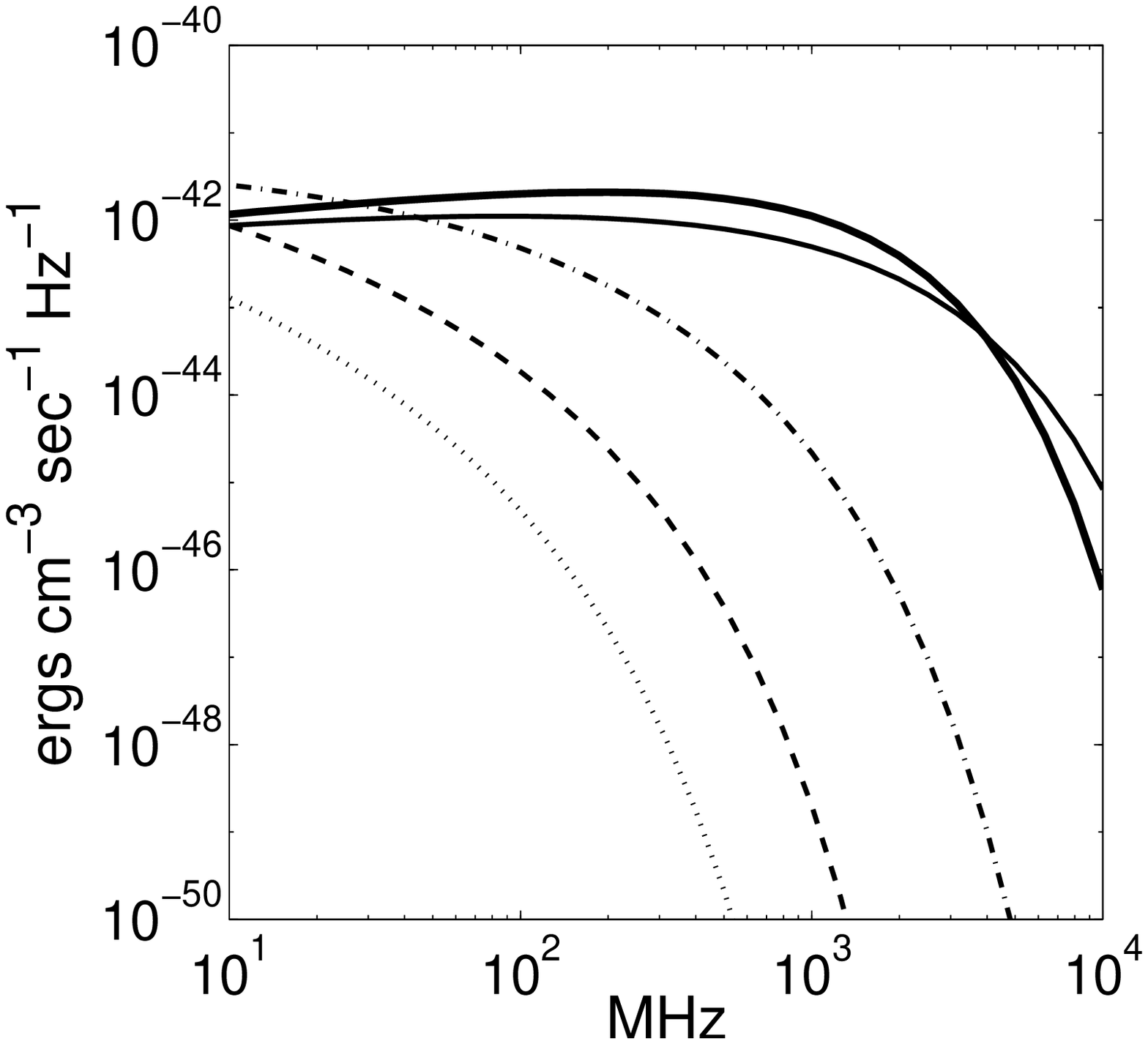}
  \caption{Radio emissivity at different radii. The parameters are from model A6. The emissivity
  at $r=$ 0 (\textit{thick solid curve}), $10\arcmin$ (\textit{solid curve}), $30\arcmin$
  (\textit{dash-dotted curve}), $50\arcmin$ (\textit{dashed curve}), and $70\arcmin$ (\textit{dotted
  curve}) are shown.
  }\label{fig9}
\end{figure}

As shown in the modelling results, a central ``plateau" in the spectral-index
distribution can be produced in clusters with a central field strength of 6
$\mu$G. To understand the origin of this characteristic feature of the radio
emission of Coma C, we use the results of Model A6 as an example and show the
electron spectra and the radio emissivity of this model at different locations
in Figures~\ref{fig8} and \ref{fig9}, respectively. It is obvious that the
electrons in the central regions are strongly accelerated to the cutoff
energies and flat spectra can be formed within the central $10\arcmin$ regions.
The ratios between the radio emissivity at 326 MHz and that at 1380 MHz are
almost equal within the central regions and, at these two frequencies, the
radio emissivity within the central regions dominates over that outside this
regions; thus a central ``plateau" may form in the spectral-index distribution.
Outside the central region, the electron spectra are not very different from
one another because the effects are mainly from the inverse Compton losses and
the spatially uniform re-acceleration, and then the differences between the
radio emissivity at different locations are mainly due to the variations of the
magnetic fields. As the radius increases, the high-frequency emissivity
gradually decreases because of weaker magnetic fields; therefore, the
progressive steepening of the spectral index is presented. Naturally, it should
also exist a decline at high frequencies in the integrated radio spectrum as
proposed by \citet{sch87}.

It is interesting to note that the size of the ``plateau" is
almost equal to the region where the magnetic field is $\gtrsim$ 3
$\mu$G (Figure~\ref{fig1}). This fact might indicate that the
formation of this ``plateau" is related to the situation that the
synchrotron losses dominate over the inverse Compton losses (e.g.,
Brunetti et al. 2001). The ``plateau" should be produced by the
combination of the radial setting of re-acceleration and that of
magnetic fields. As discussed in previous paragraph, when the
inverse Compton losses dominate, it is impossible to obtain a flat
``plateau" because the radius-dependent magnetic fields will cause
the high-frequency emissivity to decrease as the radius increase.
In other words, it is possible to obtain a flat spectral index
``plateau" in our models only when the synchrotron losses dominate
over the inverse Compton losses. If a central ``plateau" presents
in the spectral-index distribution, the central strength of the
cluster magnetic field should be greater than 3 $\mu$G. Note that
since a central ``plateau" can exist only in a period in the
evolution of radio halos, a central ``plateau" in the
spectral-index distribution may be absent even a cluster
possessing a high central field.

\begin{figure}
  \plotone{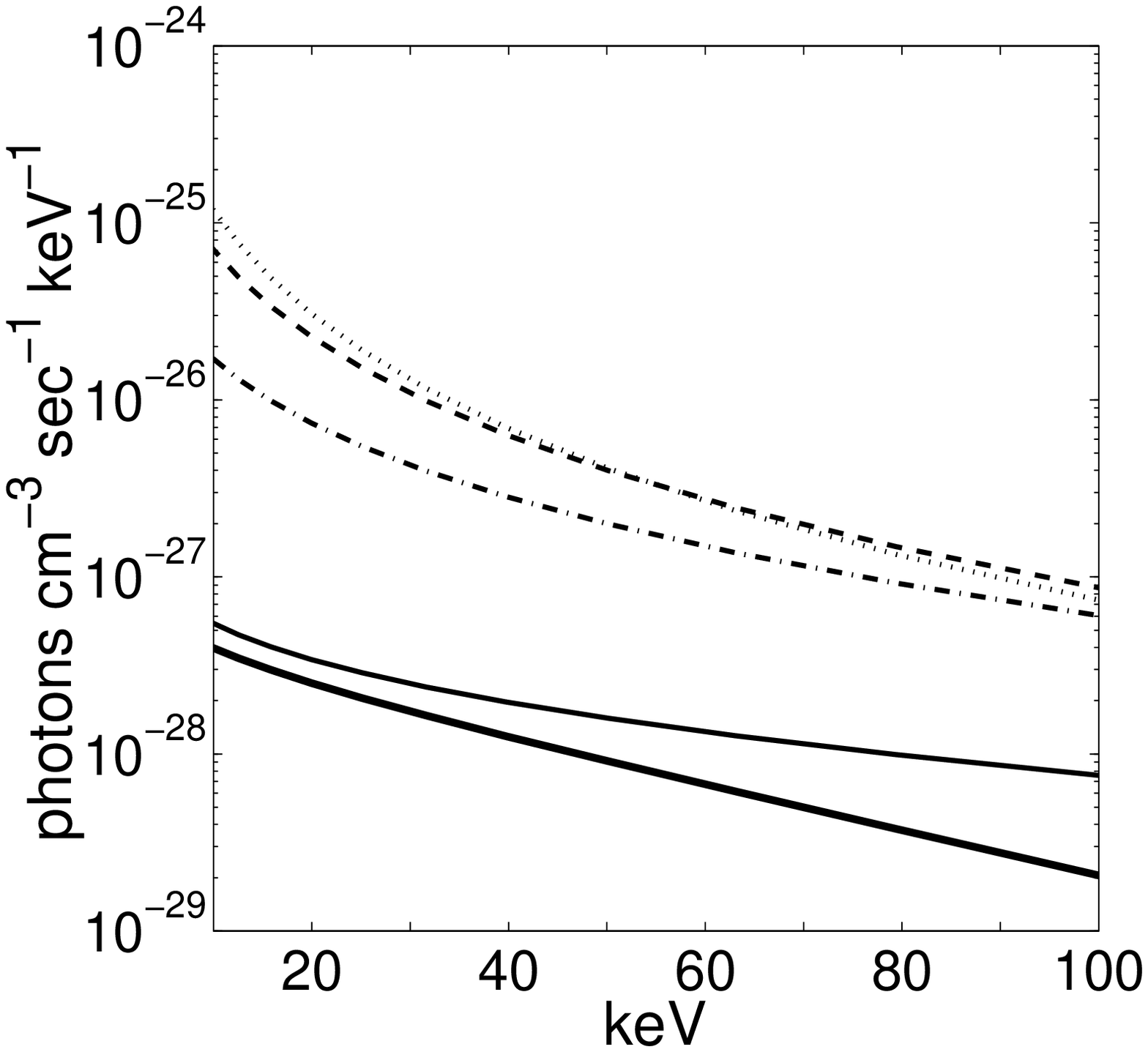}
  \caption{HXR excess emissivity at different radii. The parameters are from model A6. The emissivity at
  $r=$ 0 (\textit{thick solid curve}), $10\arcmin$ (\textit{solid curve}), $30\arcmin$
  (\textit{dash-dotted curve}), $50\arcmin$ (\textit{dashed curve}), and $70\arcmin$ (\textit{dotted
  curve}) are shown.
  }\label{fig10}
\end{figure}

The emissivity of the HXR excess of the Model A6 is shown in
Figure~\ref{fig10}. Obviously, the HXR excess is mainly
contributed from the outer regions of the cluster volume ($>
30\arcmin$) as indicated by \citet{bru01}. This might reconcile
the problem that the volume-averaged magnetic fields derived from
the HXR excesses are usually much lower than the magnetic fields
from the Faraday rotation measurements. However, we note that the
average magnetic field in the core regions adopted here is
$\lesssim$ 2 $\mu$G while the magnetic fields from the Faraday
rotation measurements are $\approx$ 6 $\mu$G. It might be possible
that a further fine-tuned version of our model can solve the
problem of the high field--ICS discrepancy.

It is expected that the profile and strength of magnetic fields
will affect our results. From the Faraday rotation measurements,
the magnetic fields are $\sim$ 4--8 $\mu$G out to $\sim$ 0.75 Mpc
from cluster centers \citep{cla01}. We note that the magnetic
fields in galaxy clusters might have a higher value and a flatter
profile in the core region than the adopted model. A higher
magnetic field requires a stronger re-acceleration to sustain the
radio emission. To produce a ``plateau" distribution of the
spectral index for a higher magnetic field with a flatter profile,
the re-acceleration function need to be almost constant over the
region. This indicates that the re-acceleration caused by the
galaxy motions might be negligible.

The ICS of the CMB photons by relativistic electrons is also suggested as the
mechanism of the extreme-ultraviolet (EUV) excess in the Coma cluster
\citep{hwa97,ens98}. However, the EUV excess possesses a narrow radial profile
and concentrates in the inner regions of the cluster \citep{bow98}, while the
HXR excess is mainly contributed from the outer regions of the cluster volume
as shown in Figure~\ref{fig9}. It is thus inadequate to assume that the
EUV-emitting electrons are also from the outside and low-magnetic-field regions
of the cluster. \citet*{tsa02} investigated the EUV excess with a $\sim 5\
\mu$G field; they showed that it is possible to explain the EUV excess and the
radio emission with such a high average magnetic field. The EUV-emitting
electrons and the radio/HXR-emitting electrons could have very different
distributions because the low-energy EUV-emitting electrons suffer much less
energy losses during their evolution and can accumulate over a much longer age.

\section{SUMMARY}

Time-independent and time-dependent re-acceleration models are studied to
investigate the formation of the radio halo and HXR excess in the Coma cluster
with a high central field strength of 6 $\mu$G. In these models, the
relativistic electrons are injected by merger shocks and re-accelerated by
ensuing violent turbulence. The effects of the strength of merger shocks on the
formation of the radio halo and HXR excess are also considered.

We have reproduced all the main features in the radio and HXR
excess emission of the Coma cluster in our models. We have
obtained a central ``plateau" in the spectral-index distribution
as observed in the Coma cluster with a high central field strength
of 6 $\mu$G in our models. The size of the ``plateau" is almost
equal to the region where the synchrotron loss dominates over the
inverse Compton loss, i.e., the magnetic field is $\gtrsim$ 3
$\mu$G. Our models can naturally produce the observed radial
steepening of the radio spectral index. We can fit the integrated
radio spectra very well, in particular, we can reproduce the
high-frequency decline feature in the spectra as measured by
\citet{thi03}. These same models can also produce the observed HXR
excess, which is mainly from the outer regions of the cluster
volume as pointed out by \citet{bru01}.

We have also shown that only merger shocks with Mach numbers in a very small
range can produce results agreeing with the observations of the radio and HXR
emission in the Coma cluster. The Mach numbers are $1.6 < \mathcal{M} < 2$,
corresponding to the initial electron power-law index $4.7 > s > 3.3$, which
are consistent with simulations and observations of merger shocks.

We have adopted a artificial magnetic field distribution in our
calculation. The real magnetic field might be higher and have a
flatter distribution than the adopted model. It is possible to
obtain similar results for a higher magnetic field with a stronger
re-acceleration. However, it might be have problems to produce the
observed HXR excess within the ICS scheme for such a high magnetic
field.

\acknowledgments

We are grateful to S. Bowyer for a careful reading of the manuscript and
several suggestions that helped to improve this paper. We thank F. Govoni for
kindly providing the data of the Coma radio profile. We are also grateful to an
anonymous referee for his valuable comments which stimulated us to
significantly improve this paper. This work was partially supported by the
National Science Council of Taiwan under NSC 91-2112-M-008-045, and by the
Ministry of Education of Taiwan through the CosPA Project 91-N-FA01-1-4-5. CYH
acknowledges support by the National Science Council through grant NSC
91-2112-M-008-039.

\clearpage
\begin{deluxetable}{cccc}
\tablecaption{Time-Independent Re-Acceleration Models \label{tbl}} \tablewidth{0pt} \tablehead{
\colhead{Model} & \colhead{t} & \colhead{$a_{0}$} &
\colhead{$a_{1}$}\\
\colhead{} & \colhead{(Gyr)} & \colhead{($10^{-16}$ s$^{-1}$)} & \colhead{($10^{-16}$ s$^{-1}$)}\\
\hline\\\multicolumn{4}{c}{$s=$ 2.5}}
\startdata
A1  &1.6 &1.87  &1.52 \\
A2  &1.8 &1.84  &1.53 \\
\hline\\
\multicolumn{4}{c}{$s=$ 3.3}\\
\hline\\
A3  &1.4 &2.18  &1.23 \\
A4  &1.6 &2.07  &1.35 \\
\hline\\
\multicolumn{4}{c}{$s=$ 4.0}\\
\hline\\
A5  &0.8 &3.40  &0.62 \\
A6  &1.0 &2.88  &0.85 \\
\hline\\
\multicolumn{4}{c}{$s=$ 4.7}\\
\hline\\
A7  &0.4 &7.05  &0.01 \\
A8  &0.6 &5.05  &0.30 \\
\enddata
\end{deluxetable}

\clearpage
\begin{deluxetable}{cccc}
\tablecaption{Time-Dependent Re-Acceleration Models with $A=1$ and $D=2$ Gyr$^{-1}$\label{tb2}}
\tablewidth{0pt} \tablehead{ \colhead{Model} & \colhead{t} & \colhead{$a_{0}$} &
\colhead{$a_{1}$}\\
\colhead{} & \colhead{(Gyr)} & \colhead{($10^{-16}$ s$^{-1}$)} & \colhead{($10^{-16}$ s$^{-1}$)}\\
\hline\\\multicolumn{4}{c}{$s=$ 3.3}}
\startdata
B1  &1.4 &1.90  &1.26 \\
B2  &1.6 &2.02  &1.16 \\
\hline\\
\multicolumn{4}{c}{$s=$ 4.0}\\
\hline\\
B3  &0.8 &2.46  &0.57 \\
B4  &1.0 &2.23  &0.86 \\
\enddata

\end{deluxetable}

\clearpage

\clearpage

\clearpage

\clearpage

\clearpage

\clearpage

\clearpage

\clearpage

\clearpage

\clearpage

\end{document}